\documentclass[aps, pra, reprint, a4paper, 10pt, superscriptaddress, nofootinbib, floatfix, amsfonts, amssymb, amsmath]{revtex4-2}
\usepackage[utf8]{inputenc}
\usepackage[T1]{fontenc}
\usepackage{braket}
\usepackage{mathtools}
\usepackage{graphicx}
\usepackage{hyperref}
\hypersetup{citecolor=magenta}
\hypersetup{colorlinks=true}
\hypersetup{linkcolor=blue}
\hypersetup{urlcolor=blue}
\newcommand\reals{\mathbb R}
\newcommand\dd{\mathrm d}

\newcommand\ee{\mathrm e}
\newcommand\PP{\mathbb P}
\newcommand\exv[1]{\langle{#1}\rangle}

\DeclareMathOperator\tr{tr}
\DeclarePairedDelimiter\abs\lvert\rvert
\newcommand\st{\text{ST}}
\newcommand\qm{\text{QM}}
\newcommand\nn{\text{neg}}
\newcommand\tn{\text{tun}}
\newcommand\bn{\text{brk}}

\begin{document}
\title{Operational Theories in Phase Space: Toy Model for the Harmonic Oscillator}

\author{Martin Pl\'{a}vala}
\email{martin.plavala@uni-siegen.de}
\affiliation{Naturwissenschaftlich-Technische Fakult\"{a}t, Universit\"{a}t Siegen, 57068 Siegen, Germany}

\author{Matthias Kleinmann}
\affiliation{Faculty of Physics, University of Duisburg–Essen, Lotharstra{\ss}e 1, 47048 Duisburg, Germany}
\affiliation{Naturwissenschaftlich-Technische Fakult\"{a}t, Universit\"{a}t Siegen, 57068 Siegen, Germany}

\begin{abstract}
We show how to construct general probabilistic theories that contain an energy observable dependent on position and momentum. The construction is in accordance with classical and quantum theory and allows for physical predictions, such as the probability distribution for position, momentum and energy. We demonstrate the construction by formulating a toy model for the harmonic oscillator that is neither classical nor quantum. The model features a discrete energy spectrum, a ground state with sharp position and momentum, an eigenstate with non-positive Wigner function as well as a state that has tunneling properties. The toy model demonstrates that operational theories can be a viable alternative approach for formulating physical theories.
\end{abstract}

\maketitle

\emph{Introduction.}---%
Various ideas have been proposed to generalize quantum theory. For example, quaternionic \cite{Adler-quaternionicQT} and non-hermitian \cite{Bagarello-nonHermitian, BenderBrodyMuller-nonHermitian} reformulations of quantum theory were investigated, as well as a more general, operational, approaches to physical theories. While the former are closely enough related to quantum theory to allow for experimental tests \cite{ProcopioRozemaWongZiHamelOBrienZhangDakic-quaternionic}, the operational approaches studied to date are bound to black-box-like, device-independent and finite-dimensional proto-theories, which do not describe actual physical systems, but rather information-theoretic effects, like, for example, bounds on violations of Bell inequalities \cite{PohJoshiSiddarthCereCabelloKurtseifer-photonExperiments,MazurekPuseyReschSpekkens-experimentalTomography,WeilenmannColbeck-selfTesting}. The operational approaches we have in mind here, are built on the assumption that convexity represents mixtures of states and are collected under the term general probabilistic theories \cite{Mueller-GPTreview}. Within this framework, for example, thermodynamics \cite{ChiribellaScandolo-thermodynamics, KrummBarnumBarrettMuller-thermodynamics}, different notions of entropy \cite{KimuraIshiguroFukui-entropy, BarnumBarrettClarkOrloffLeiferSpekkensStepanikWilceWilke-entropy}, dynamics \cite{GrossMullerColbeckDahlsten-dynamics, BranfordDahlstenGarner-hamiltonDynamics}, and recently even the operational consequences of gravitational effects \cite{GalleyGiacominiSelby-gravity} were investigated.

\begin{figure}
\includegraphics[width=\linewidth]{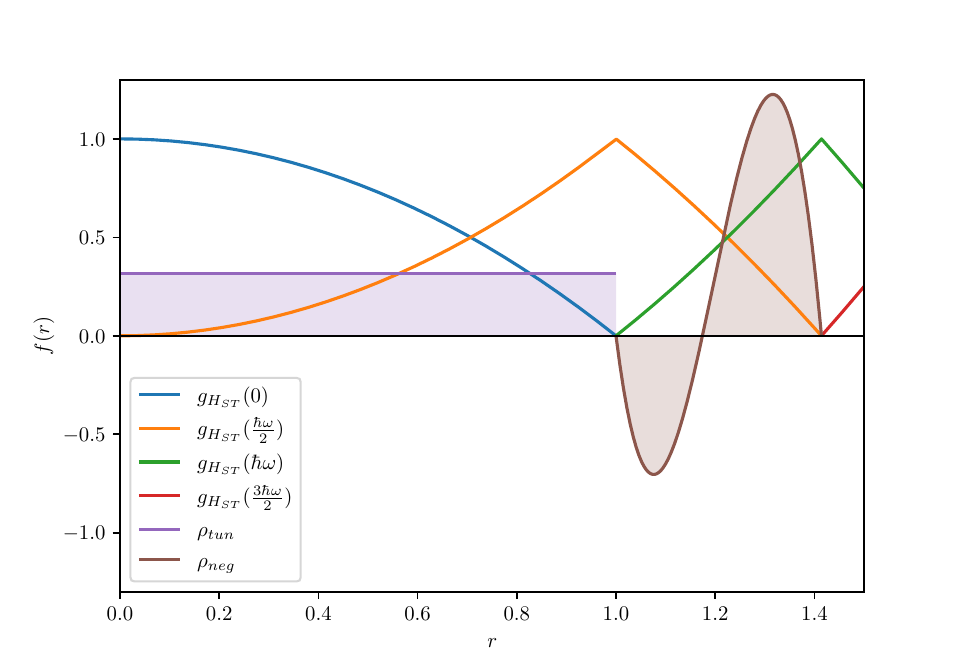}
\caption{Phase space spectral measure and phase space states for the sawtooth oscillator as a function of $r=\sqrt{(p^2/\hbar m\omega)+(m\omega q^2/\hbar)}$ in units of $\hbar$. The functions of the phase space spectral measure for the energies $E_0 = 0$, $E_1 = \frac{\hbar \omega}{2}$, $E_2 = \hbar \omega$, and $E_3 = \frac{3 \hbar \omega}{2}$ are shown in blue, orange, green, and red, respectively. The ``tunneling'' state $\rho_\tn$ and the eigenstate $\rho_\nn$ for the energy $E_2$ are shown in purple and brown, respectively, with the area under the functions filled.}
\label{fig:intro}
\label{fig:ST-tun}
\label{fig:ST-neg}
\end{figure}

The goal of this Letter is to offer a bridge between operational theories in the above sense and extensions of quantum theory mentioned earlier. We accomplish our goal by constructing an operational theory where the energy is linked to position and momentum, analogically to classical and quantum theory. Thus we are able to formulate a toy model for the most archetypal of all physical system---the harmonic oscillator. Our construction demonstrates that there are alternatives to quantization and operator formalism for building physical theories and that one can use the operational approach for such a construction.

Generalized theories with continuous position and momentum were investigated before \cite{Spekkens-epistricted}. In these theories, states are described by pseudo-probability densities, that is by real-valued and possibly nonpositive functions $\rho(q,p)$ such that $\int_{\reals} \rho(q,p) \dd p$ is the probability density for the random variable $\tilde q$ corresponding to a position measurement and similarly for momentum. For a function $f(q,p)$, one obtains the mean value of $f$ via $\exv{ \tilde{f} }_\rho = \int_{\reals^2} f(q,p) \rho(q,p) \dd q \dd p$. This approach is consistent with classical theory and with the Wigner function formalism of quantum theory \cite{Wigner-WignerFunctions,Groenewold-QM,Moyal-WignerFunctions,Rosen-WignerFunctions,Cohen-WignerFunctions,Bergeron-CTtoQT,HilleryConnellScullyWigner-WignerFunctions,Fairlie-wignerFunctions}. Hence, we can compute the mean value $\exv{\tilde H}_\rho$ of the energy, but we cannot compute the probability distribution of the energy, or, equivalently, we cannot compute its higher moments $\exv{\tilde{H}^k}_\rho$. But it is crucial for any theory to allow us to compute the probability distribution of $\tilde H$ since using the probability distribution we can, for example, determine whether a state is an eigenstate of the energy observable and, even more important, we can determine the spectrum of the energy observable. This is no small feat, as predicting the spectrum of the hydrogen atom was one of the first results of quantum theory and to this day finding the energy spectrum of various Hamiltonians is an important problem.

The first naive solution would be to compute $\exv{\tilde{H}^2}_\rho$ as the mean value of $H^2(q,p)$, but even in quantum theory we have $\exv{\tilde{H}^2}_\rho \neq \int_{\reals^2} H^2(q,p) \rho(q,p) \dd q \dd p$, see \cite{Case-wignerFunctions}. One can even show that $\exv{\tilde{H}^k}_\rho = \int_{\reals^2} H^k(q,p) \rho(q,p) \dd q \dd p$ only holds in classical theory \cite{PlavalaKleinmann-WGPTmath}. Another naive solution would be to treat energy as an independent variable $\epsilon$ and to have pseudo-probability densities of the form $\rho(q,p,\epsilon)$. But then energy is not linked to position and momentum and so we do not follow this approach.

We solve this problem by introducing phase space spectral measures. With these measures we achieve our goal to obtain the probability distribution of $\tilde{H}$ and to describe the energy spectrum of a general system in a similar way to using the spectral measure of an operator in quantum theory. In fact one can express both quantum and classical theories using phase space spectral measures, that is, our general construction includes both theories as special cases. Thus one can also use our construction as a groundwork for finding axioms that would uniquely specify quantum theory among other theories, which would generalize the known results for finite-dimensional systems \cite{Hardy-derivationQT,MasanesMuller-derivatonQT,ChiribellaDArianoPerinotti-derivationQT,Wilce-derivationQT}.

As a demonstration of generality of our construction, we present the toy model of the sawtooth oscillator; it is an infinite-dimensional general probabilistic theory with continuous position and momentum and with the energy observable expressed using position and momentum, as in classical and quantum theory. But the sawtooth oscillator is neither classical, nor quantum, nor any transitional form of classical and quantum oscillators. The sawtooth oscillator is characterized by the sawtooth-shaped phase space spectral measure $g_{H_\st}$ and it has an eigenstate $\rho_\nn$ that is given by a nonpositive pseudo-probability density. We also show that the model exhibits tunneling properties for an appropriately chosen state $\rho_\tn$. Both states, as well as the spectral measure are depicted in Fig.~\ref{fig:intro}.

\emph{Phase space spectral measures in operational theories.}---%
We work with a general operational theory on phase space where states are given as pseudo-probability densities $\rho(q,p)$ and observables are real-valued phase-space functions $A(q,p)$. The states are hence real-valued phase space functions with normalization $\int_{\reals^2} \rho(q,p) \dd q \dd p = 1$. Importantly, the function $\rho(q,p)$ is not required to be a probability density and hence may attain negative values for some regions in phase space. The mean value of the random variable $\tilde A$ associated to the outcomes of a measurement of the observable $A$ is given by the phase-space integral
\begin{equation}
\label{eq:mean-ps}
\exv{\tilde A}_\rho = \int_{\reals^2} A(q,p)\rho(q,p) \dd q \dd p.
\end{equation}

As outlined in the introduction, this formalism is not yet sufficient to describe the probability distribution of $\tilde A$. To enable a full probabilistic description, we require that each observable $A$ has associated a phase space spectral measure $g_A$ with the property that the probability for $\tilde A$ to attain a value in the set $I\subset \reals$ is given by
\begin{equation}
\PP( \tilde A\in I ) = \int_{\reals^2} g_A(I;q,p) \rho(q,p) \dd q\dd p.
\end{equation}
Consequently, this measure is a pseudo-probability measure at each phase space point, $g_A(q,p)\colon \reals \supset I \mapsto g_A(I;q,p)$. That is, $g_A(\emptyset;q,p) = 0$, $g_A(\reals;q,p) = 1$, and $g_A(q,p)$ is countably additive on disjoint sets, meaning that for a countable collection $(I_n)_n$ of mutually disjoint subsets of $\reals$, we have $g_A(\cup_n I_n;q,p) = \sum_n g_A(I_n;q,p)$. We have $\PP_{\rho}(\tilde A \in \reals) = 1$ from the normalization of $g_A$ and $\rho$ and for the mean value of $\tilde A$ we get
\begin{equation}
\begin{split}
\exv{\tilde{A}}_\rho &= \int_\reals a\, \PP_{\rho}(\tilde A = a) \, \dd a \\
&= \int_\reals \int_{\reals^2} a g_A(I;q,p) \rho(q,p) \, \dd q \dd p \dd a.
\end{split}
\end{equation}
To make this equation to coincide with Eq.~\eqref{eq:mean-ps} we impose the consistency condition
\begin{equation}
\label{eq:construct-gSum}
A(q,p)= \int_\reals a g_A(a;q,p) \, \dd a.
\end{equation}

In order for a given $\rho$ to be a phase-space density, we require that the probability densities of position and momentum are given as the marginals $\int_\reals \rho(q,p) \dd p$ and $\int_\reals \rho(q,p) \dd p$, respectively. This allows us identify the phase space spectral measure of the position observable $q$ as
\begin{equation}
\label{eq:construct-positionPSSM}
g_{q}(I;q,p)=\int_I \delta(q-\xi) \, \dd \xi,
\end{equation}
and analogously for the momentum observable $p$.

At this point we mention a generic way to construct a phase space spectral measure: Let $(t_n)_n$, $t_n\colon \reals\to\reals$, be a family of functions and $(a_n)_n$ a corresponding family of eigenvalues such that $\sum_n t_n(x) = 1$ and $\sum_n a_n t_n(x) = x$. Then a phase space spectral measure for $A$ is given by
\begin{equation}
\label{eq:tconst}
g_A(I;q,p) = \sum_{\tau a_n \in I} t_n[A(q,p) / \tau],
\end{equation}
where $\tau$ is a constant with the same units as $A$.

\emph{Phase space spectral measures in quantum theory.}---%
We illustrate now how phase-space spectral measures are obtained in the phase-space formulation of quantum theory. We start with a short review of the Wigner--Weyl formalism, for a full review see \cite{Case-wignerFunctions}. One defines the Weyl transform of a self-adjoint operator $\hat A$ as
\begin{equation}
\label{eq:weyl}
\hat{A}^W(q,p) = \int_{\reals} \ee^{-\frac i\hbar p \cdot y} \braket{q + \frac{y}{2} | \hat{A} | q - \frac{y}{2} } \, \dd y,
\end{equation}
where $\ket{x}$ denotes the formal eigenvector of the position operator $\hat q$ with eigenvalue $x$. The density operator is represented in phase space using the Wigner transform
\begin{equation}
\label{eq:wigner}
\rho^W (q,p) = h^{-1} \int_{\reals} \ee^{-\frac i\hbar p \cdot y} \braket{
q+\frac{y}{2} | \rho | q-\frac{y}{2} } \, \dd y.
\end{equation}
Note, that both, Eq.~\eqref{eq:weyl} and Eq.~\eqref{eq:wigner}, describe the same transformation up to the factor of $h^{-1}$ and both transformations yield real-valued functions. The Wigner transform of a state is normalized and can attain negative values, but its marginals are the probability density of position and momentum. In the general case of an arbitrary observable $\hat A$, its mean value is obtained as
\begin{equation}
\label{eq:quantum-meanValue}
\int_{\reals^2} \hat A^W(q,p) \rho^W(q,p) \, \dd q\dd p = \tr(\hat A\rho)= \exv{\tilde A}_\rho,
\end{equation}
where $\tilde A$ again denotes the random variable corresponding to a measurement of $\hat A$.

In order to obtain the phase space spectral measure of an observable we use that the probability of observing a value in a given set of values $I$ is given by $\PP_\rho(\tilde{A} \in I) = \tr(\rho \Pi^{A}_I)$, where $\Pi^{A}_I$ is the spectral measure of $\hat{A}$, that is, $\hat{A} = \int_\reals a \Pi_a^{A} \dd a$. This probability $\tr(\rho \Pi^{A}_I)$ can also be seen as the mean value of the operator $\Pi^{A}_I$ and so we can use Eq.~\eqref{eq:quantum-meanValue} to express this mean value in terms of functions on phase space; the Weyl transform of $\Pi^{A}_I$ yields the phase space spectral measure $g_{A}(I;q,p)$, that is, $g_{A}(I;q,p) = \int_I (\Pi_a^{A})^W(q,p)\,\dd a$.

To illustrate this construction, assume that $\hat{A}$ has discrete and non-degenerate spectrum with eigenvalues $a_n$ and $\rho^W_n$ the Wigner functions of the corresponding eigenstates. Then
\begin{equation}
g_{A}(I;q,p) = \sum_{a_n \in I} h \rho^W_n(q,p).
\end{equation}
Note that one can also define phase space spectral measures in classical theory, see Appendix~\ref{appendix:classical}.

\emph{Time-evolution and positivity in operational theories.}---%
For the purpose of this Letter we assume that that the time-evolution in an operational theory is given by the Liouville equation $\dot{\rho} = \{ H, \rho \}$, with $\{f,g\}=\frac{\partial f}{\partial q}\frac{\partial g}{\partial p}-\frac{\partial f}{\partial p}\frac{\partial g}{\partial q}$ the Poisson bracket and where $H(q,p)$ is the energy observable. In general one can also use other possible Hamiltonians as generators of other translations, but for simplicity we will consider only the time-translations.

Note, that in the Wigner--Weyl formalism, the Poisson bracket is replaced by the Moyal bracket \cite{Moyal-WignerFunctions, Fairlie-wignerFunctions}. The Moyal bracket contains quantum corrections of the order $\hbar^2$ and higher, but it is equal to the Poisson bracket for simple Hamiltonians, such as for the harmonic oscillator. Yet, in a general theory, one could imagine a different dynamical equation, but we choose here an equation that reproduces the situation for the quantum and classical harmonic oscillator.

In order to get a consistent theory we must require that all observable probabilities are positive. Thus if $A(q,p)$ is an observable in our theory with phase space spectral measure $g_A(I;q,p)$, then we must have
\begin{equation}
\label{eq:construct-positivity}
\PP_{\rho}(\tilde A\in I) \geq 0, \text{ for all } I \subset \reals.
\end{equation}
Naively one would say that any pseudo-probability density $\rho(q,p)$ that satisfies positivity for all observables should be a valid state. This is not the case, because we also have to require that the time-evolution preserves the positivity. This condition is nontrivial, see Appendix~\ref{appendix:broken} for an example.

Note that the positivity conditions for the position and momentum observables during time-evolution are closely related to any linear combination of position and momentum observables being a well-defined observable. This property holds in both classical and quantum theory and is used as a defining property for Wigner representations on discrete phase spaces \cite{Wootters-discreteWigner, GibbonsHoffmanWootters-discreteWigner, Gross-discreteWigner, DeBrotaStacey-discreteWigner, SchwonnekWerner-discreteWigner}.

We define the set of states using the no-restriction hypothesis \cite{ChiribellaDArianoPerinotti-GPTpurification,FilippovGudderheinnosaariLeppajarvi-restrictions} as the largest set of pseudo-probability densities that satisfies the positivity condition for all (future) times. Since the left-hand-side of Eq.~\eqref{eq:construct-positivity} and the time-evolution are linear in $\rho$, it follows that the set of states is convex. Thus the resulting theory is a general probabilistic theory and the dimension of the theory is infinite, because already the spectral measures for $q$ and $p$ contain infinitely many linearly independent functions, for example, $g_q( I_n; q,p)$ where $I_n$ is any open interval $(n,n+1)$.

\emph{The sawtooth oscillator.}---%
The quantum and classical harmonic oscillator have both the same phase space representation $\hat H^W=H_c= \frac{p^2}{2m}+\frac{m\omega^2}2 q^2$ for the energy observable and in both models the time-evolution is given by the Liouville equation, see Appendix~\ref{appendix:quantum} for a short review of the Wigner--Weyl formalism for the quantum harmonic oscillator. The significant difference between both models lies in the phase space spectral measure and the corresponding eigenstates for the energy observable.

We now introduce the sawtooth oscillator as a toy model which follows the same principles but is neither classical nor quantum. That is, the energy coincides with the classical case, $H_\st = H_c$ and position and momentum have the phase space spectral measures given as in Eq.~\eqref{eq:construct-positionPSSM}. Our sawtooth model is defined according to Eq.~\eqref{eq:tconst} with the family of functions $t_n=T_n$, $n\ge 0$, the values $a_n=n$ and $\tau=\frac12\hbar\omega$. The function $T_n$ are depicted in Fig.~\ref{fig:ST-Tn} and defined in Appendix~\ref{appendix:sawtoothPSSM}. Hence, our phase space spectral measure reads
\begin{equation}
g_{H_\st}(I; q,p) = \sum_{n=0}^\infty \Big( \int_I \delta[ \hbar \omega \frac n2 - \epsilon ] \dd \epsilon \Big) T_n (r^2),
\end{equation}
where $r^2= 2H_c(q,p)/\hbar\omega$. By construction, the energy spectrum of the sawtooth oscillator is discrete, with energies $\hbar \omega \frac n{2}$ for $n = 0,1,2, \dotsc$ From $\sum_n T_n(x)=1$ it follows that the normalization condition is satisfied, $g_{H_\st}(\reals; q,p) = \sum_{n=0}^\infty T_n (r^2) = 1$ and furthermore, due to $\sum_n n T_n(x)=x$, Eq.~\eqref{eq:construct-gSum} is also satisfied,
\begin{equation}
H_\st(q,p) = \sum_{n=0}^\infty \hbar \omega\frac{n}{2} T_n (r^2) = \frac12 \hbar\omega r^2= H_c(q,p).
\end{equation}

\begin{figure}
\includegraphics[width=\linewidth]{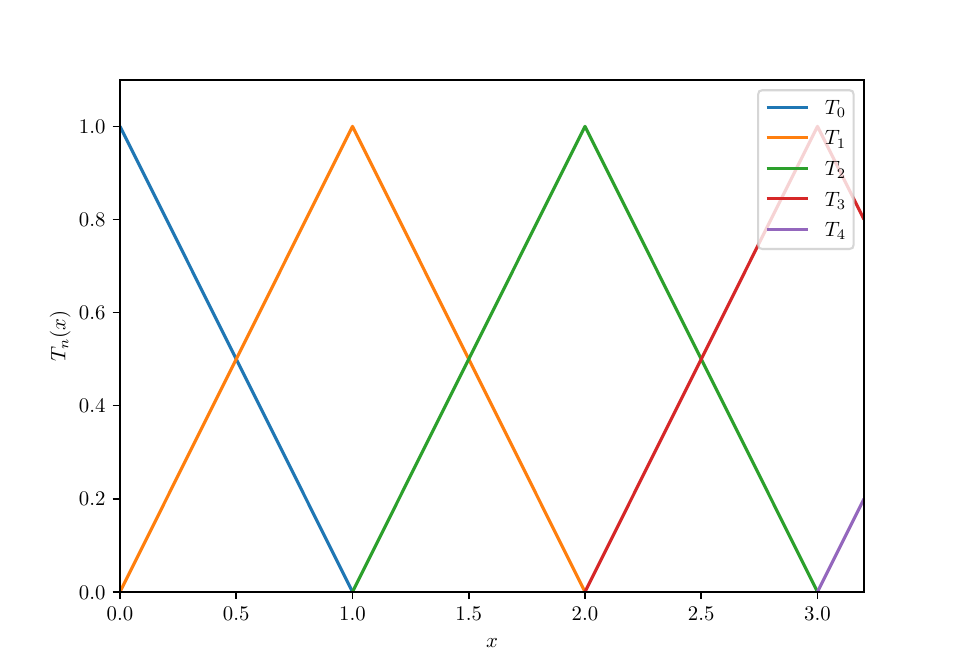}
\caption{The functions $T_n$ used in the construction of the phase space spectral measure for the energy observable of the sawtooth oscillator.} \label{fig:ST-Tn}
\end{figure}

We mention that, following our general construction scheme, one can in principle apply similar sawtooth constructions to other systems, for example the hydrogen atom with $H_\mathrm{H}(q,p)=\frac{\abs{p}^2}{2m}-\frac\kappa{\abs q}$. By extending $T_0$ and $T_1$ to negative $x$, $T_0(x)=1-x$ and $T_1(x)=x$, we only have to adapt the constant $\tau$, for example, $\tau=-m\kappa^2/2\hbar^2$. Then, Eq.~\eqref{eq:tconst} yields a phase space spectral measure for $H_\mathrm{H}$ with spectrum $0,\tau, 2\tau, \ldots$ However, this construction is rather naive, for example, the spectrum does not have a lower bound and the construction completely ignores the role of the angular momentum.

Returning to the sawtooth oscillator, it is different from the classical and quantum case as can be demonstrated by considering possible states in the model. First, we consider the state $\rho_0(q,p) = \delta(q)\delta(p)$. Then we have $ \PP_{\rho_0} ( \tilde{H} = 0 ) = \exv{g_{H_\st}(0), \rho_0} = 1 $ and so $\rho_0$ is a time-invariant eigenstate of the sawtooth oscillator corresponding to zero energy. This state is completely localized in the phase space, that is, the preparation uncertainty of both position and momentum is jointly zero and hence is at variance with quantum theory. Second, the sawtooth oscillator is not classical: For $\rho_\nn$ the nonpositive pseudo-probability density as in Fig.~\ref{fig:ST-neg} and defined in Appendix~\ref{appendix:nonpositive}, one can show that $\rho_\nn$ is an eigenstate of the sawtooth oscillator corresponding to the energy $\hbar \omega$. The nonpositivity of $\rho_\nn$ implies that we cannot jointly measure the position and momentum of $\rho_\nn$, hence the state demonstrates nonclassicality in the toy model. We also define the ``tunneling'' state $\rho_\tn$, see Fig.~\ref{fig:ST-tun} and Appendix~\ref{appendix:tunneling}. This state is not an eigenstate of the sawtooth oscillator, it has probability $\frac12$ for $\tilde H=0$ and $\tilde H= \tfrac{\hbar \omega}{2}$. We discuss now in which sense this state has tunneling behavior.

\emph{Quantum tunneling in the sawtooth oscillator.}---%
We use the definition of tunneling presented in Ref.~\cite{LinDahlsten-tunneling}. Usually one says that a quantum particle is tunneling, if the particle crosses a potential barrier that is higher than the energy of the particle. This definition is not applicable to the harmonic oscillator because the potential is not in the form of a barrier. But in the standard scenario, if the particle is able tunnel through a potential barrier, there must be a nonzero probability of observing the particle inside the potential barrier, that is, the wave function of the particle must penetrate into the barrier. One can generalize this statement as follows. The probability of observing a particle in the region inside the barrier is higher than the probability of the particle having energy higher than the energy of the barrier. In this sense a state $\rho$ has tunneling behavior if there is some threshold $\alpha$ such that
\begin{equation}
\label{eq:tunneling}
\PP_\rho [ V(\tilde q) > \alpha ] > \PP_\rho [ \tilde{H}>\alpha]
\end{equation}
where $V(q)$ is the potential energy and $\tilde{H}$ is again the random variable corresponding to the total energy. According to the classical intuition, one would expect that the potential energy is upper bounded by the total energy of the particle. This does not have to be the case in general, even in the standard formulation of tunneling the particle has non-zero probability of being localized inside the barrier, which is a region where the potential energy is higher than the total energy of the particle. Eq.~\eqref{eq:tunneling} formalizes this using the respective probabilities. When this definition is applied to the quantum harmonic oscillator one finds \cite{LinDahlsten-tunneling} that the wave function of the ground state has tunneling behavior, which in this case means that the wave function of the ground state spreads more than its energy would allow according to classical intuition.

Returning to the sawtooth oscillator, for the state $\rho_0$, we have $\PP_{\rho_0}[ \tilde{H} = 0 ] = 1$ but $\PP_{\rho_0} [ V(\tilde q) > 0] = 0$ and therefore $\rho_0$ has no tunneling behavior. Also, for the nonclassical eigenstate $\rho_\nn$ we find $\PP_{\rho_\nn} [ \tilde{H} = \hbar \omega] = 1$ but $\PP_{\rho_\nn} [ V(q) > \hbar \omega ] = 0$. It follows that the eigenstate $\rho_\nn$ with nonzero energy does not exhibits tunneling behavior, contrary to the quantum case.

In contrast, the state $\rho_\tn$ has a positive probability density and so one may expect that it must behave according to classical intuition, but this is not true. Because the phase space spectral measure $g_{H_\st}$ is different from the classical one, positivity of $\rho_\tn$ does not imply classicality. For $0 < \alpha < \frac{\hbar \omega}{2}$ we have $ \PP_{\rho_\tn} [ \tilde{H} >\alpha] = \tfrac{1}{2}. $ Since $V(q) = \frac{1}{2} m \omega^2 q^2$, it follows that $\PP_{\rho_\tn} [ V(\tilde{q})>0]=1-\PP_{\rho_\tn}[\tilde{q}^2=0]=1$ and so for a sufficiently small $\alpha > 0$ we must have $ \PP_{\rho_\tn} [ V(\tilde q) > \alpha] > \tfrac{1}{2}$, see Appendix~\ref{appendix:tunneling} for details. Hence $\rho_\tn$ exhibits tunneling behavior for a sufficiently small threshold $\alpha$.

\emph{Conclusions.}---%
The main conceptual result of this Letter is that the constructed phase space spectral measures are key to an operational approach for working with energy and other observables that depend on position and momentum. Using this result, we have essentially formulated all possible theories of the harmonic oscillator, up to introducing exotic dynamics beyond what we observe in classical and quantum theory. One can clearly extend the construction to the case of multiple particles by adding additional pairs of variables $q_i, p_i$. Using this approach, one can construct a field theory with exotic properties, since the sawtooth oscillator has discrete energy spectrum, but the ground state has zero energy. In general, the phase space spectral measure of the energy does not have to have an equidistant distribution of the energy levels, therefore one can obtain a field theory without well-defined photons, which may give rise to the prediction of new physical effects.

As already pointed out, one can also formulate theory of hydrogen atom analogically to the sawtooth oscillator, but the results are not fully satisfying. Apart from the aforementioned problems with the energy spectrum and lack of angular momentum, one also has to consider the time-evolution, since in quantum mechanics the time-evolution of the hydrogen atom is no longer given by the Liouville equation. Hence one needs to find some generalization of the Schr\"{o}dinger equation and some other phase space spectral measure with more physical spectrum, but then one also needs to verify whether these choices are consistent and whether they produce an operational theory with satisfactory physical properties, for example the phase space spectral measure should be stationary. Further discussion of these problems is left for future work.


\acknowledgments
\emph{Acknowledgments.}---%
We thank J. Siewert for discussions. This work was supported by the Deutsche Forschungsgemeinschaft (DFG, German Research Foundation, project numbers 447948357 and 440958198), the Sino-German Center for Research Promotion (Project M-0294) and the ERC (Consolidator Grant 683107/TempoQ). MP acknowledges support from the Alexander von Humboldt Foundation.

\bibliography{citations.bib}



\onecolumngrid

\appendix

\section{Phase space spectral measures of classical harmonic oscillator} \label{appendix:classical}
In classical mechanics the situation is very similar to the standard formalism in statistical mechanics. The state of a system is described by a probability density $\rho_c(q,p)$ in phase space and observables are represented by phase space functions $A_c(q,p)$, such that $ \exv{\tilde A}_{\rho_c}= \int_{\reals^{2n}} A_c(q,p) \rho_c(q,p) \dd^n q\dd^n p$. Due to the probability density $\rho_c$, the phase space function $A_c$ also gives rise to a random variable $\tilde A$. The probability to obtain a value in a set $I\subset \reals$ is given by
\begin{equation} \label{eq:classical-Aprobability}
\PP_{\rho_c}(\tilde A\in I) = \int_{\reals^{2n}}\int_{I} \delta(A_c(q,p)-a) \rho_c(q,p) \, \dd a \dd^n q\dd^n p.
\end{equation}
This allows us to identify the classical analog of the phase space spectral measure
\begin{equation}
g_{A_c}(I;q,p)=\int_I\delta(A_c(q,p)-a) \, \dd a.
\end{equation}

\section{Wigner--Weyl formalism for the quantum harmonic oscillator} \label{appendix:quantum}
We provide a short review of the quantum harmonic oscillator with the classical Hamilton function $H_c(q,p) = \frac{p^2}{2m} + \frac{1}{2} m\omega^2 q^2$. The Weyl transform of the corresponding Hamilton operator $\hat H$ yields back $H_c(q,p)$, that is, $\hat H^W(q,p) = H_c(q,p)$. The Wigner functions of the $n$-th eigenstate with energy $\hbar\omega(n+\frac12)$ are well-known to be
\begin{equation}
\rho_n^W= h^{-1}\,2(-1)^n \ee^{-r^2} L_n(2 r^2),
\end{equation}
where $L_n$ are the Laguerre polynomials and $r^2= 2H_c(q,p)/\hbar\omega$. Using the relation between the Wigner transform and the Weyl transform we obtain the phase space spectral measure
\begin{equation}
g_{\hat H}(I)= \sum_{n=0}^\infty\Big(\int_{I} \delta(\hbar\omega(n+\tfrac12)-\epsilon) \,\dd \epsilon\Big) h\rho_n^W,
\end{equation}
In particular, this yields $g_{\hat H}(\reals)=\sum_n h\rho_n^W\equiv 1$ and
\begin{equation}
\begin{split}
H_\qm(q,p) &= \int_\reals \epsilon' g_{\hat H}(\epsilon';q,p)\dd\epsilon' = H_c(q,p).
\end{split}
\end{equation}
In the case of the harmonic oscillator, the time-evolution of the state $\rho^W$ given by the von-Neumann equation simplifies to the time-evolution given by the Liouville equation. For general Hamiltonians such a simple relation does not hold and one has to use the Moyal bracket.

\section{The phase space spectral measure of the sawtooth oscillator} \label{appendix:sawtoothPSSM}
The functions $T_n$ shown in Fig.~\ref{fig:ST-Tn} in the main text are given as
\begin{equation}
\label{eq:Tn-def}
T_n(x) =
\begin{cases}
x - (n-1) & x \in [n-1, n] \\ -x + n+1 & x \in [n, n+1] \\ 0 & x \notin [n-1, n+1],
\end{cases}
\end{equation}
where $n$ is a nonnegative integer. Note that for $x \geq 0$ we have $T_n(x) \neq 0$ only if $n$ is the floor or the ceiling of $x$, i.e., when $n = \lfloor x \rfloor$ or $n = \lceil x \rceil$. Assume for simplicity that $0 \leq x \leq 1$, we then have
\begin{equation}
\sum_{n=0}^\infty T_n(x) = T_0(x) + T_1(x) = -x+1 + x = 1.
\end{equation}
It then follows that the same holds for all $x \geq 0$ because the function $T_n$ just periodically repeat each other, i.e., we have
\begin{equation}
\label{eq:Tn-k}
T_n(x+k) = T_{n-k}(x)
\end{equation}
for $x \geq 0$ and $n,k \in \mathbb{N}$, $n \geq k$. It follows that we have
\begin{equation}
\sum_{n=0}^\infty T_n(x) = 1
\end{equation}
for all $x \geq 0$. Similarly, let again $0 \leq x \leq 1$, then we have
\begin{equation}
\sum_{n=0}^\infty n T_n(x) = T_1(x) = x.
\end{equation}
One can again use \eqref{eq:Tn-k} to get
\begin{equation}
\sum_{n=0}^\infty n T_n(x) = x
\end{equation}
for all $x \geq 0$.

\section{Pseudo-probability density broken by time-evolution} \label{appendix:broken}
\begin{figure}
\includegraphics[width=0.75\linewidth]{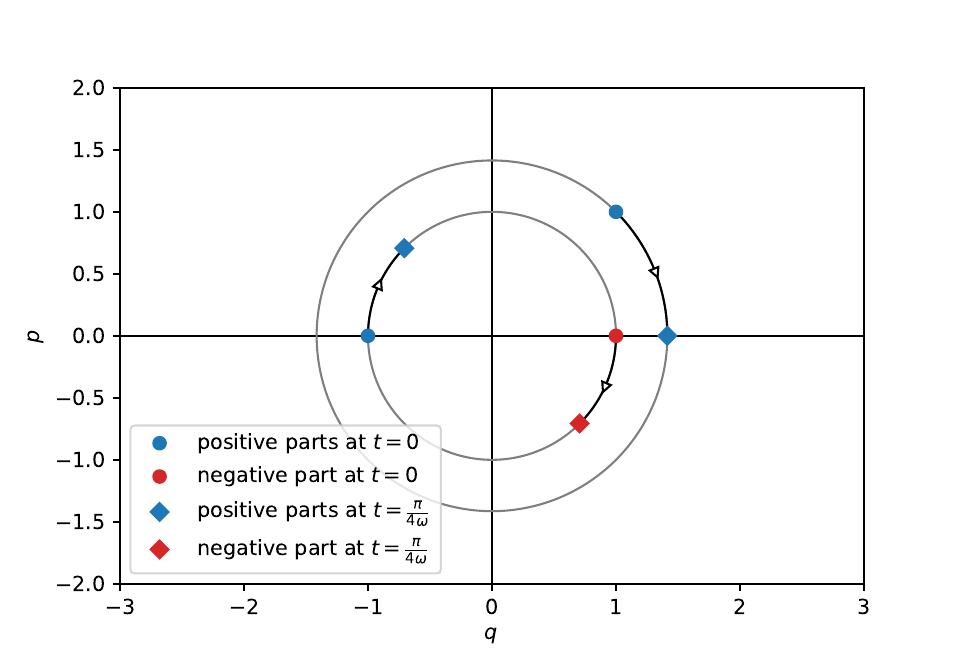}
\caption{The positive and negative parts of the states $\rho_\bn = \rho_\bn(t=0)$ and $\rho_\bn(t=\frac{\pi}{4 \omega})$, plotted for $\hbar = \omega = m = 1$. The black lines with arrows indicate how the different points move during the time-evolution from $t = 0$ to $t = \frac{\pi}{4 \omega}$, and the gray circles are the trajectories in phase space along which the points would travel during time-evolution. One can clearly see that for $t = 0$ the marginals of $\rho_\bn$ are positive probability densities, because the negative contribution is countered by a positive contribution. For $t = \frac{\pi}{4 \omega}$ this is not the case anymore, hence $\rho_\bn$ is not a valid state of the sawtooth oscillator.}
\label{fig:brk}
\end{figure}

We will present an example of a pseudo-probability density $\rho_\bn$ that satisfies the conditions in Eq.~\eqref{eq:construct-positivity} in the main text for time $t = 0$, but not for $t > 0$. Let
\begin{equation}
\label{eq:rhobn-def}
\rho_\bn(q,p) = \delta \left( q + \sqrt{\dfrac{\hbar}{m \omega}} \right) \delta(p) - \delta \left( q - \sqrt{\dfrac{\hbar}{m \omega}} \right) \delta(p) + \delta \left( q - \sqrt{\dfrac{\hbar}{m \omega}} \right) \delta \left( p - \sqrt{\hbar \omega m} \right),
\end{equation}
see also Fig.~\ref{fig:brk} for a graphical representation of $\rho_\bn$. It is straightforward to verify that $\rho_\bn$ is properly normalized, i.e., that we have
\begin{equation}
\int_{\reals^2} \rho_\bn(q,p) \, \dd q \dd p = 1.
\end{equation}
We will now proceed to show that the positivity conditions
\begin{align}
\int_I \int_\reals \rho(q,p) \, \dd q \dd p \geq 0, \quad\text{and} \quad
\int_\reals \int_J \rho(q,p) \, \dd q \dd p \geq 0, \label{eq:ST-marginals} \\
\int_{\reals^2} g_{H_\st}(I; q,p) \rho(q,p) \, \dd q \dd p \geq 0, \label{eq:ST-gPos}
\end{align}
hold for $\rho_\bn$. Here $g_{H_\st}$ is the phase space spectral measure of the sawtooth oscillator introduced in the main text. To show that conditions in Eq.~\eqref{eq:ST-marginals} hold, we compute the marginals
\begin{align}
\int_\reals \rho_\bn (q,p) \, \dd p &= \delta \left( q + \sqrt{\dfrac{\hbar}{m \omega}} \right), \\
\int_\reals \rho_\bn (q,p) \, \dd q &= \delta \left( p - \sqrt{\hbar \omega m} \right),
\end{align}
which are both positive. To verify the condition in Eq.~\eqref{eq:ST-gPos}, note that we have
\begin{equation}
\int_{\reals^2} g_{H_\st} \left( \dfrac{\hbar \omega n}{2} ; q,p \right) \rho_\bn(q,p) \, \dd q \dd p = \int_{\reals^2} T_n (r^2) \rho_\bn(q,p) \, \dd q \dd p = T_n (2) \geq 0
\end{equation}
because
\begin{align}
r^2 \left(\pm \sqrt{\dfrac{\hbar}{m \omega}}, 0 \right) &= 1, \\ r^2 \left(\sqrt{\dfrac{\hbar}{m \omega}}, \sqrt{\hbar \omega m} \right) &= 2,
\end{align}
where $r^2 = r^2(q,p) = H_c(q,p)/\frac12\hbar\omega$.

To get the time-evolution for the harmonic oscillator, one can solve the Hamilton equations and get
\begin{align}
q(t) &= \dfrac{p_0}{m \omega} \sin(\omega t) + q_0 \cos(\omega t), \\ p(t) &= p_0 \cos(\omega t) - q_0 m \omega \sin(\omega t),
\end{align}
where $(q_0, p_0)$ is the initial position and momentum of the system at $t = 0$. We then have
\begin{equation}
\rho(q_0,p_0;t) = \rho(q(t), p(t)),
\end{equation}
where $q_0 = q(0)$ and $p_0 = p(0)$. Let $t = \frac{\pi}{4 \omega}$, then
\begin{align}
\rho_\bn \left( q,p;\dfrac{\pi}{4 \omega} \right) &=
\delta \left( \dfrac{q + \dfrac{p}{m \omega}}{\sqrt{2}} + \sqrt{\dfrac{\hbar}{m \omega}} \right) \delta \left(\dfrac{p - m \omega q}{\sqrt{2}} \right)
- \delta \left( \dfrac{q + \dfrac{p}{m \omega}}{\sqrt{2}} - \sqrt{\dfrac{\hbar}{m \omega}} \right) \delta \left(\dfrac{p - m \omega q}{\sqrt{2}} \right) \\
&+ \delta \left( \dfrac{q + \dfrac{p}{m \omega}}{\sqrt{2}} - \sqrt{\dfrac{\hbar}{m \omega}} \right) \delta \left( \dfrac{p - m \omega q}{\sqrt{2}} - \sqrt{\hbar \omega m} \right),
\end{align}
see Fig.~\ref{fig:brk} for a graphical representation. By integrating over $p$ we get
\begin{equation}
\int_\reals \rho_\bn \left( q,p;\dfrac{\pi}{4 \omega} \right) \dd p =
\delta\left( q + \sqrt{\dfrac{2 \hbar}{m \omega}} \right)
+ \delta(q)
- \delta\left( q - \sqrt{\dfrac{2 \hbar}{m \omega}} \right)
\end{equation}
which is not positive, and so for sufficiently small $\varepsilon > 0$ we get
\begin{equation}
\PP_{\rho_\bn \left(\frac{\pi}{4 \omega}\right)} \left( \tilde{q} \in \left[ \sqrt{\dfrac{2 \hbar}{m \omega}} - \varepsilon, \sqrt{\dfrac{2 \hbar}{m \omega}} + \varepsilon \right] \right) = -1 < 0
\end{equation}
which violates the first condition in Eq.~\eqref{eq:ST-marginals}. One can also see this from Fig.~\ref{fig:brk}, because integrating over $p$ simply means that we project all of the three parts of $\rho_\bn(t)$ onto the $q$ axis. For $t = 0$ the negative part is projected to the same point as one of the positive parts and they cancel out, but for $t = \frac{\pi}{4 \omega}$ this is not the case anymore. Note that the time $t = \frac{\pi}{4 \omega}$ was chosen just to simplify the calculations, but it follows from Fig.~\ref{fig:brk} that $\rho_\bn(t)$ has nonpositive marginals for any $\omega t \notin \{0, \frac{\pi}{2} + 2k \pi, (2k+1) \pi, \frac{3\pi}{2} + 2k \pi, 2 k \pi \}$ for $k \in \mathbb{N}$.

\section{The nonpositive eigenstate of the sawtooth oscillator} \label{appendix:nonpositive}
The state $\rho_\nn$ is given for $1 \leq r \leq \sqrt{2}$ as
\begin{equation}
\label{eq:rhonn-def}
\rho_\nn(q,p) = \dfrac{30(1-r)\Big(8(147\sqrt{2} - 208) + 2(738\sqrt{2} - 1043)r + 7(65\sqrt{2} - 92)r^2\Big)}{(2827\sqrt{2} - 3998) \pi \hbar}
\end{equation}
and $\rho_\nn(q,p) = 0$ everywhere else, see also Fig.~\ref{fig:intro} in the main text. One can easily see that for $n = 0$ and for $n \geq 3$ we have
\begin{equation}
\PP_{\rho_\nn} \left( \tilde{H}_\st = \dfrac{n \hbar \omega}{2} \right) = \int_{\reals^2} g_{H_\st} \left( \dfrac{n \hbar \omega}{2}; q,p \right) \rho_\nn(q,p) \, \dd q \dd p = 0.
\end{equation}
The state $\rho_\nn$ was found using Mathematica to be a third-order polynomial in $r$ such that
\begin{align}
\PP_{\rho_\nn} \left( \tilde{H}_\st = \dfrac{\hbar \omega}{2} \right) &= 0 \\
\PP_{\rho_\nn} \left( \tilde{H}_\st = \hbar \omega \right) &= 1,
\end{align}
i.e., so that $\rho_\nn$ is an eigenstate corresponding to the energy $\hbar \omega$.

One can verify that the conditions in Eq.~\eqref{eq:construct-positivity} in the main text are satisfied for $\rho_\nn$. We will not provide explicit formulas, since they are lengthy. Note that we have
\begin{equation}
\{ (q,p) : V(q) > \hbar \omega \} \subset \{ (q,p) : H_\st(q,p) > \hbar \omega
\} = \{ (q,p) : r(q,p) > \sqrt{2} \}.
\end{equation}
and it follows that
\begin{equation}
\PP(V(\tilde q)>\hbar\omega)=\PP_{\rho_\nn} ( \tilde{q} \in \{ q: V(q) > \hbar
\omega \}) \leq \PP_{\rho_\nn} ( \tilde{r} > \sqrt{2} )
\end{equation}
where $\tilde{r}$ is the random variable corresponding to measuring $r$ as a function of energy. Note that $\PP_{\rho_\nn} ( \tilde r > \sqrt{2} ) = 0$ follows directly from \eqref{eq:rhonn-def}, but it is also clear from Fig.~\ref{fig:intro} in the main text.

\section{The ``tunneling'' state of the sawtooth oscillator} \label{appendix:tunneling}
Let $r = r(q,p)$ be given by $r^2 = H_c(q,p)/\frac12\hbar\omega$. The probability density $\rho_\tn$ is given as
\begin{equation}
\label{eq:rhotn-def}
\rho_\tn(q,p) =
\begin{cases}
\dfrac{1}{\pi \hbar} & r \leq 1 \\
0 & r > 1.
\end{cases}
\end{equation}
This function defines a classical state of the sawtooth oscillator. To verify that $\rho_\tn$ is properly normalized, note that $\dd q \dd p = \hbar r \dd r \dd \varphi$ and we have
\begin{equation}
\int_{\reals^2} \rho(q,p) \, \dd q \dd p = 2 \int_0^1 r \, \dd r = 1.
\end{equation}
For $n \geq 2$, the supports of $\rho_\tn$ and $g_{H_\st}(\frac{n \hbar \omega}{2})$ are disjoint, as seen in Fig.~1 in the main text, so we have
\begin{equation}
\PP_{\rho_\tn} \left( \tilde{H}_\st = \dfrac{n \hbar \omega}{2} \right) = \int_{\reals^2} g_{H_\st} \left( \dfrac{n \hbar \omega}{2}, q,p \right) \rho_\tn(q,p) \, \dd q \dd p = 0.
\end{equation}
For $n = 0$ we have
\begin{equation}
\PP_{\rho_\tn} ( \tilde{H}_\st = 0 ) = \int_{\reals^2} T_0(r^2) \rho_\tn (q,p) \, \dd q \dd p = 2 \int_0^1 (-r^2 + 1) r \, \dd r = \dfrac{1}{2}.
\end{equation}
For $n = 1$ we have
\begin{equation}
\PP_{\rho_\tn} \left( \tilde{H}_\st = \dfrac{\hbar \omega}{2} \right) = \int_{\reals^2} T_1(r^2) \rho_\tn (q,p) \, \dd q \dd p = 2 \int_0^1 r^3 \, \dd r = \dfrac{1}{2}.
\end{equation}
The probability density of position of $\rho_\tn$ is
\begin{equation}
\int_\reals \rho_\tn (q,p) \, \dd p =
\begin{cases}
\dfrac{2 m \omega}{\pi \hbar} \sqrt{ \dfrac{\hbar}{m \omega} - q^2} & |q| \leq \sqrt{\dfrac{\hbar}{m \omega}} \\
0 & |q| > \sqrt{\dfrac{\hbar}{m \omega}}
\end{cases}
\end{equation}
and we then have
\begin{equation}
\PP_{\rho_\tn} (\tilde{q} \in I) = \dfrac{2 m \omega}{\pi \hbar} \int_{I \cap Q} \sqrt{ \dfrac{\hbar}{m \omega} - q^2} \, \dd q,
\end{equation}
where
\begin{equation}
Q = \left\lbrace q : |q| \leq \sqrt{\dfrac{\hbar}{m \omega}} \right\rbrace.
\end{equation}
Let $0 < \alpha < \frac{\hbar \omega}{2}$, we then have
\begin{equation}
\PP_{\rho_\tn} ( V(\tilde q)> \alpha ) = \dfrac{2 m \omega}{\pi \hbar}
\,2\int\limits_{ \sqrt{\frac{2 \alpha}{m \omega^2}}}^{\sqrt{\frac{\hbar}{m \omega}}} \sqrt{ \dfrac{\hbar}{m \omega} - q^2} \, \dd q = 1 - \dfrac{2}{\pi}
\left( \arcsin\left( \sqrt{\dfrac{2 \alpha}{\hbar \omega}} \right) +
\sqrt{\dfrac{2 \alpha}{\hbar \omega}} \sqrt{1 - \dfrac{2 \alpha}{\hbar \omega}}
\right).
\end{equation}
Now consider that $\alpha \ll \frac{\hbar \omega}{2}$, then up to the first order in $\sqrt{\alpha}$ we have
\begin{equation}
\PP_{\rho_\tn} ( V(\tilde q) > \alpha ) \approx 1 - \dfrac{4}{\pi}
\sqrt{\dfrac{2 \alpha}{\hbar \omega}}
\end{equation}
and so choosing $\alpha = \frac{\hbar \omega}{2} \frac{\pi^2}{16} \varepsilon^2$ we get
\begin{equation}
\label{eq:rhotn-varepsilon}
\PP_{\rho_\tn} ( V(\tilde q) > \alpha ) \approx 1 - \varepsilon.
\end{equation}

\end{document}